\documentclass[debug,overfull,cite,figures]{epl} 
\makeatletter 
\def\@pnumwidth{2em} 
\makeatother 
\usepackage{graphicx}
\usepackage{dcolumn}
\usepackage{bm}
\def\xx{\tilde{x}_2} 
\def\xxm{\tilde{x}_{\rm 2min}} 
 
\title{Conductance distribution in 3D Anderson insulators: 
deviation from log-normal form}

\shorttitle{Conductance distribution in 3D Anderson insulators} 
\author{%
P. Marko\v{s}\inst{1}\thanks{E-mail: \email{markos@savba.sk}} \and 
K. A. Muttalib\inst{2} \and 
P. W\"olfle\inst{3} \and 
J. R. Klauder\inst{2}\thanks{Also at Department of Mathematics.} 
} 
\shortauthor{P. Marko\v{s} \etal} 
\institute{ 
\inst{1} Institute of Physics, Slovak Academy of Sciences - 845 11 
Bratislava, Slovakia\\ 
\inst{2} Department of Physics, University of Florida, 
Gainesville, FL 32611-8440\\ 
\inst{3} Institut f\"ur Theorie der Kondensierten Materie, 
Universit\"at Karlsruhe, Germany} 
 

\begin{document} 
\maketitle

\begin{abstract} 
{We show how a recent proposal to obtain the distribution of 
conductances in three dimensions (3D) from a generalized 
Fokker-Planck equation for the joint probability distribution of 
the transmission eigenvalues can be implemented for all strengths 
of disorder by numerically evaluating certain correlations of 
transfer matrices. We then use this method to obtain analytically, 
for the first time, the 3D conductance distribution in the 
insulating regime and provide a simple understanding of why it differs 
qualitatively from the log-normal 
distribution of a quasi one-dimensional wire.} 
\end{abstract} 
 
It has been shown recently that for a quasi one-dimensional (Q1D) 
disordered quantum wire, the distribution of conductances $P(g)$ 
(at zero temperature, in the absence of electron-electron 
interaction) has many surprising features arising from large 
mesoscopic fluctuations. These include a highly asymmetric 
`one-sided' log-normal distribution at intermediate disorder 
between the metallic and insulating limits 
\cite{mu-wo-99,go-mu-wo-02}, and a singularity in the distribution 
near the dimensionless conductance $g\sim 1$ in the insulating 
regime \cite{mu-wo-ga-go-03}. Although there is no phase 
transition in Q1D conductors, numerical studies support the 
conjecture that some of these features persist in higher 
dimensions as well \cite{markos1,markos2}, which may have 
important consequences for the Anderson metal-insulator transition 
in three dimensions (3D) \cite{lee}. However, while a systematic 
method has recently been developed to obtain analytically the full 
$P(g)$ in Q1D \cite{mu-wo-go-03}, no such method is yet available 
in higher dimensions. Indeed, it has not been possible so far to 
study analytically even the simpler case of the distribution 
$P(g)$ for a 3D insulator \cite{comment,noteA2}. 
 
A phenomenological 
generalization of the Q1D method to study the distribution of 
conductances in higher dimensions has been proposed recently 
\cite{mu-go-02}. For a given dimensionality, the generalization 
involves an unknown matrix $K$ to be determined from numerical 
studies of the properties of transmission matrices for conductors 
of various strengths of disorder. Within a set of well-defined 
approximations, this method is applicable in principle for all 
strengths of disorder in all dimensions. In the present work we 
first of all test these approximations and show that they are 
valid in 3D. We then study the matrix $K$ in 3D and show that it 
can be modeled reasonably well by a single parameter in both the 
metallic and insulating limits, and that it contains information 
about the critical regime. We find that while the value of the 
parameter in the metallic regime in 3D remains the same as in Q1D, 
it changes dramatically across the transition and in the 
insulating regime. This implies that while the 3D metal is similar 
to a Q1D metal, the localization in  3D  is qualitatively 
different from the localization in Q1D, discussed in Refs.  [1-3]. 
We use the value of the parameter in the insulating limit to 
obtain analytically, for the first time, the full $P(g)$ of a 3D 
insulator. We show e.g. that the density of Lyapunov exponents in 
3D is qualitatively different from that in Q1D; while the 
conductance in the insulating regime is dominated by the smallest Lyapunov 
exponent in Q1D, this is no longer true for 3D. The 
resulting 3D $P(g)$ is \textit{not} log-normal even in the deeply 
insulating regime. Our results are in 
agreement with direct numerical studies.

The distribution of conductances in Q1D was obtained within 
the transfer matrix approach. In this approach, a conductor of 
length $L_z$ and width $L\times L$ is placed between two perfect 
leads; the scattering states at the Fermi energy then define 
$N\propto L^2$ channels. The $2N \times 2N$ transfer matrix $M$ 
relates the flux amplitudes on the right of the system to those on 
the left \cite{muttalib}. Flux conservation and time reversal 
symmetry (we consider the case of unbroken time reversal symmetry 
only) restricts the number of independent parameters of $M$ to 
$N(2N+1)$ and $M$ can be written in general as \cite{dmpk} 
\begin{equation}\label{one} 
M=\left(\matrix{ u & 0  \cr 0 & {u}^* \cr }\right) \left(\matrix{ 
\sqrt{1+\lambda} & \sqrt{\lambda}   \cr \sqrt{\lambda}   & 
\sqrt{1+\lambda} \cr }\right)\left(\matrix{ v & 0  \cr 0 & {v}^* 
\cr }\right), 
\end{equation} 
where $u,v$ are $N \times N$ unitary matrices, and $\lambda$ is a 
diagonal matrix, with positive elements $\lambda_i, i=1,2, ...N$. 
The restriction of the approach to Q1D arises from the 
``isotropy'' approximation, that the distribution $p_{L_z}(M)$ is 
independent of the matrices $u$ and $v$. In \cite{mu-go-02}, this 
restriction was lifted in favor of an unknown phenomenological 
matrix 
\begin{equation}\label{two} 
K_{ab}\equiv \left<k_{ab}\right>_L; \;\;\;  k_{ab} \equiv 
\sum_{\alpha}^N|v_{a\alpha}|^2|v_{b\alpha}|^2, 
\end{equation} 
where the angular bracket represents an ensemble average 
\cite{note1}. In terms of this matrix, the distribution 
$p_{L_z}(M)$ satisfies an evolution equation given by 
\cite{mu-go-02} 
\begin{equation} 
\label{three} 
\frac{\partial p_{L_z}(\lambda)}{\partial (L_z/l)} 
=\frac{1}{\bar{J}}\sum_a^N 
\frac{\partial}{\partial\lambda_a}\left[\lambda_a(1+\lambda_a)K_{aa} 
\bar{J}\frac{\partial p}{\partial \lambda_a}\right],~~~ 
\bar{J}\equiv\prod_{a<b}^N|\lambda_a-\lambda_b|^{\gamma_{ab}}; 
\;\;\; \gamma_{ab}\equiv\frac{2K_{ab}}{K_{aa}}. 
\end{equation} 
 
In Q1D, the matrix $K$ reduces to 
$K_{ab}=\frac{1+\delta_{ab}}{N+1}$, which implies $\gamma_{ab}=1$, 
and one recovers the well known DMPK equation with the symmetry 
parameter 
 $\beta=1$ \cite{dmpk}. 
In 3D, $K$ is not known analytically. We obtain it from direct 
numerical studies of a tight binding Anderson model defined by the 
Hamiltonian 
\begin{equation}\label{AndHam} 
{\cal H}=W\sum_n \varepsilon_nc_n^\dag c_n+\sum_{[nn']} 
t_{nn'}c^\dag_n c_{n'}. 
\end{equation} 
In (\ref{AndHam}), $n=(xyz)$ counts sites on the lattice of the 
size $L\times L\times L_z$, and $\varepsilon_n$ are random 
energies, uniformly distributed in the interval 
$[-\frac{1}{2},\frac{1}{2}]$. The parameter $W$ measures the 
strength of the disorder. The Fermi energy is chosen as 
$E_F=0.01$. The hopping term $t_{nn'}$ between the 
nearest-neighbor sites $nn'$ is unity for hopping along the $z$ 
direction and $t_{nn'}=t$ for hopping in the $x$ and $y$ 
directions.  To avoid closed channels existing in perfect leads, 
we use $t=0.4$. Then the model (\ref{AndHam}) exhibits a 
metal-insulator transition at $W_c\approx 9$. In order to 
construct a simple model for the matrix $K$, we use the method of 
\cite{pendry} to calculate numerically the matrix $TT^\dag$ ($T$ 
is the transmission matrix). Using Eq.~(\ref{one}), we have 
$TT^\dag=v^*(1+\lambda)^{-1}v$. Diagonalization of $TT^\dag$ gives 
us both $\lambda$ and all elements of the matrix $v$.

\begin{figure}
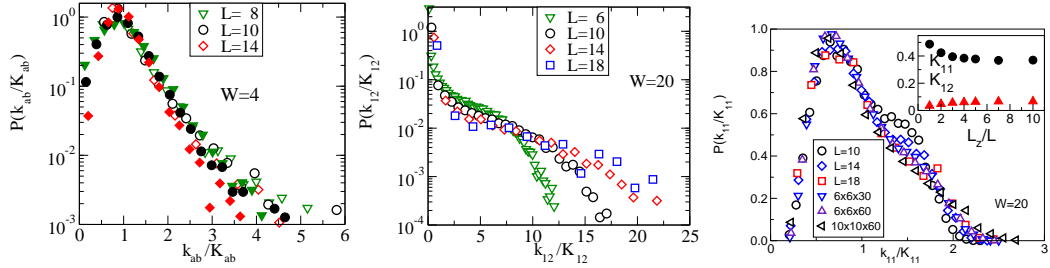
 
\centerline{%
\includegraphics[clip,width=0.22\textheight]{mmwk-fig1a.eps} 
\includegraphics[clip,width=0.22\textheight]{mmwk-fig1b.eps} 
\includegraphics[clip,width=0.22\textheight]{mmwk-fig1c.eps} 
} 
\caption{Left: the distributions of \textit{normalized} parameters 
$k_{11}$ (open symbols) and $k_{12}$ (full symbols) 
in the metallic regime ($W=4$). 
Two right figures show the same for 
the insulator ($W=20$). 
Inset of right figure shows $K_{11}$ and $K_{12}$ 
of  systems $6\times 
6\times L_z$ as a function of $L_z$.} 
 
\label{tws} 
\end{figure}

There are two major assumptions made in \cite{mu-go-02} in 
deriving Eq.~(\ref{three}), (i) the elements $k_{ab}$ can be 
replaced by their mean values $K_{ab}$ and (ii) the 
$L_z$-dependence of $K_{ab}$ is negligible. To test these 
assumptions, we analyzed the probability distributions $P(k_{ab}/K_{ab})$ 
in the insulating regime and compared them with those for metals 
(figure \ref{tws}). 
In the metallic regime all distributions seem to be  self-averaging 
(they become narrower when $L$ increases), with sharp maximum at 1. 
In the insulating regime 
$P(k_{12}/K_{12})$ 
has a very sharp distribution too, 
but  possesses a  long exponential  tail to values $k_{12}\gg K_{12}$. 
$P(k_{11}/K_{11})$ is broader, with variance of the same order as the mean. 
This means that while assumption (i) remains 
valid to leading order, fluctuations in the diagonal elements $k_{aa}$ in 
the 
insulating regime can be important if the final results are sensitive to 
the exact values of the parameters. We will therefore concentrate on the 
qualitative features of the distribution of conductances, 
which are 
insensitive to those fluctuations. 
Figure 
\ref{tws} also shows that $K_{ab}$ depend slightly on the ratio 
$L_z/L$  and reach $L_z$-independent limiting values when 
$L_z/L\to\infty$. This is qualitatively consistent with assumption 
(ii) and agrees with previous numerical analysis of parameters 
$x_i$ ($\lambda_i\equiv\sinh^2 x_i$) \cite{pichard,footn1}. We 
therefore conclude that at least to leading approximation, 
the generalized 
DMPK equation (\ref{three}) can be used in 3D at all disorder 
strengths.

\begin{figure} 
\centerline{\includegraphics[clip,width=0.22\textheight]{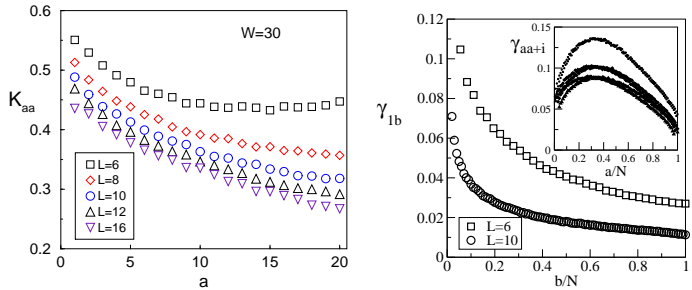} 
\includegraphics[clip,width=0.22\textheight]{mmwk-fig2b.eps}} 
\caption{Left: $a$-dependence of $K_{aa}$ for $W=30$. Right: 
$\gamma_{1b}$ for two values of $L$ and (in inset) $\gamma_{aa+i}$ for 
$i=1,2$ and 3 ($W=30$, $L=10$). 
} 
\label{w20_kaa} 
\end{figure}

\begin{figure} 
\centerline{\includegraphics[clip,width=0.22\textheight]{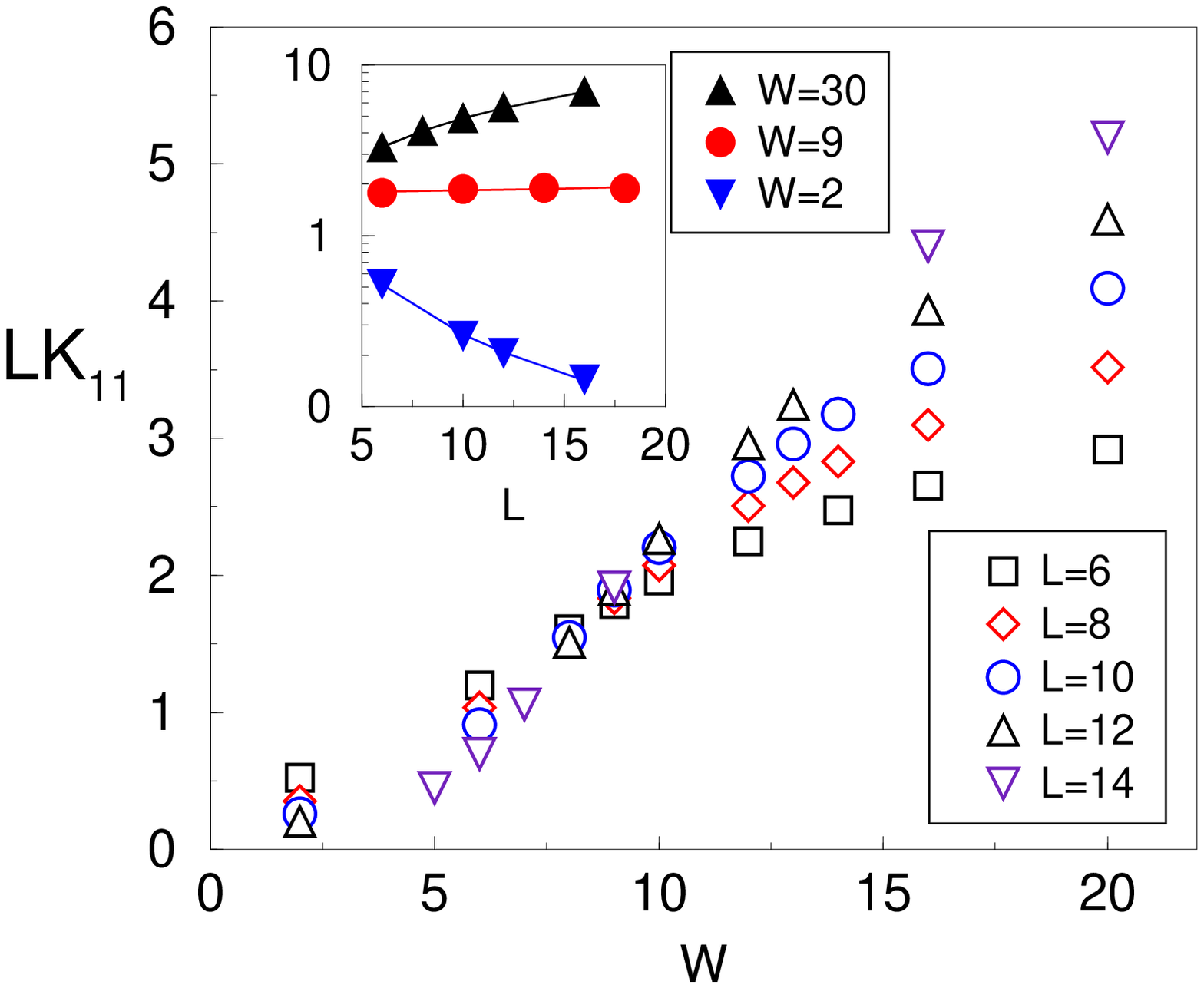} 
\includegraphics[clip,width=0.22\textheight]{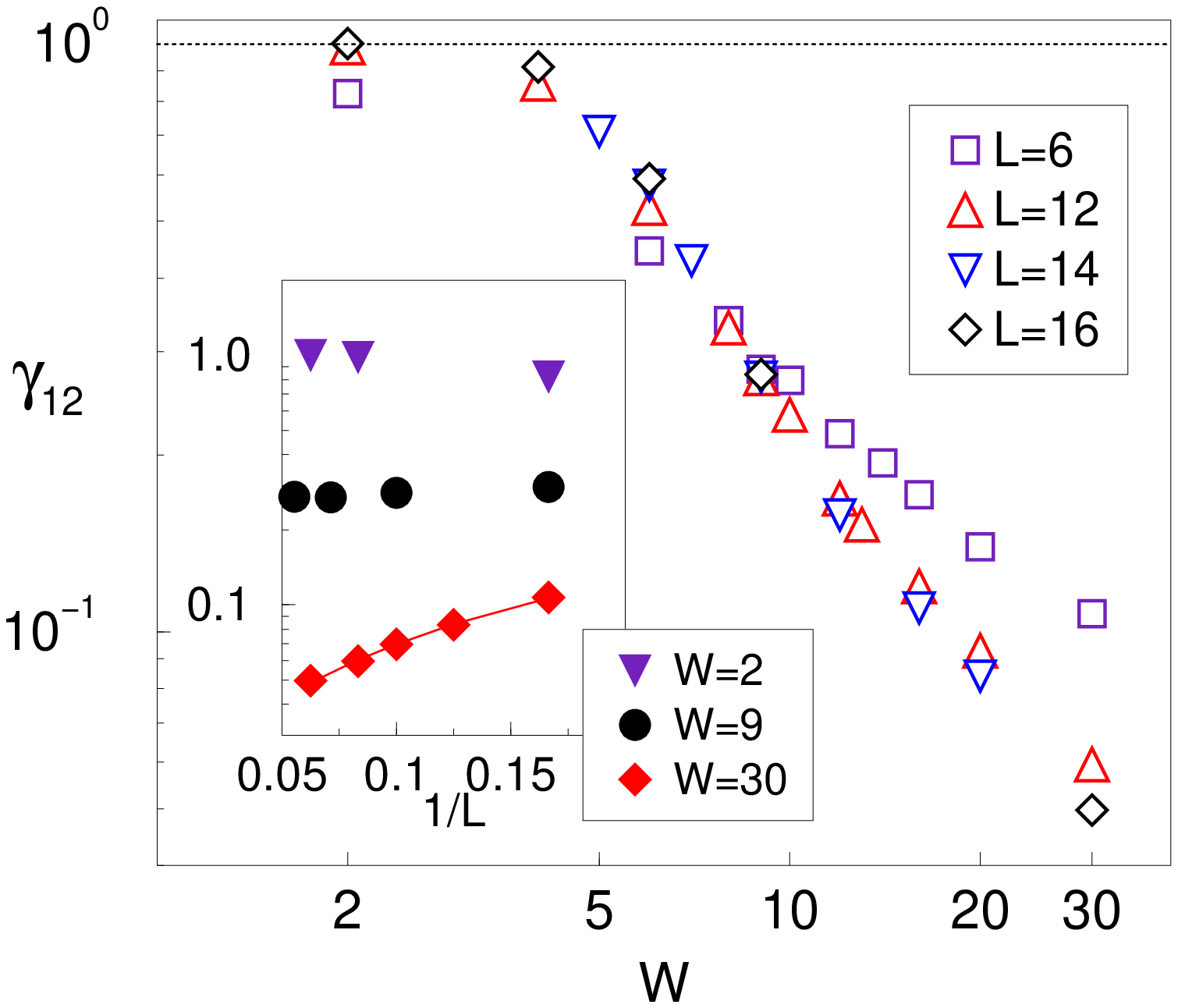} 
\includegraphics[clip,width=0.20\textheight]{mmwk-fig3c.eps}} 
\caption{Left: $LK_{11}$ for various $L$ as a function of 
disorder. Note the common crossing point for $W\approx 9$, showing 
$K_{11}\propto 1/L$ at the critical point. Inset shows the 
$L$-dependence of $LK_{11}$ for three values of disorder $W$. 
Solid lines are power fits $LK_{11}\propto L^m$ with $m=-1.3, 0.06$ and 
0.8 for for $W=2$ (metal), $W=9$ (critical point) and $W=30$ 
(insulator), respectively. 
 The linear fit for $W=30$ uses 
$K_{11}=0.37+1.2/L$. Middle:  $\gamma_{12}$ as a function of 
disorder for various system sizes. In the metallic regime, 
$\gamma_{12}\approx 1$. At the critical point, $\gamma_{12}\approx 
0.25$. In the localized regime, $\gamma_{12}$ decreases with 
increasing $W$. Inset shows  $\gamma_{12}(L)$ 
for $W=2,9$ and $30$. Solid line is a fit 
$\gamma_{12}=0.015+0.56/L$. Right figure confirms that $\gamma_{12}\propto 
1/L$ in the localized regime. Limiting values of $\gamma_{12}(L\to\infty)$ are 
given in the legend. 
} 
\label{K_11} 
\end{figure}

We now construct a model for the matrix $K$ in the insulating 
regime. Figure \ref{w20_kaa} shows $K_{aa}$ as a function of $a$ 
and $\gamma_{ab}$ for different $(a,b)$ in this regime. While 
there is some dependence on the indices for both $K$ and $\gamma$, 
the dependence is weak compared to the dependence of the matrix 
elements on disorder in the $L\rightarrow \infty$ limit. As only 
few channels contribute to the conductance 
in the insulating regime, 
we will ignore this index dependence and use $K_{aa}\approx K_{11}$ and 
$\gamma_{ab}\approx \gamma_{12}$ \cite{note}. Figure \ref{K_11} is 
consistent 
with $K_{11}\propto 1/L^m$, where $m=2,1,0$ in the metallic, 
critical and insulating limits, respectively, in agreement with 
\cite{chalker}. 
Figure \ref{K_11} also shows $\gamma_{12}$ as a function of 
disorder. It is unity in the weakly disordered metallic regime, 
showing the same behavior as in the Q1D limit. This is true for 
all $\gamma_{ab}$, which implies that a 3D metal is very similar 
to a Q1D metal. However, as the disorder is increased for given 
$L$, the value of $\gamma_{12}$ starts to decrease. Several plots 
with different $L$ values show that in the insulating limit 
$\gamma_{12}\propto 1/L$ (right figure \ref{K_11}). Thus our data show that the behavior of 
\textit{strongly disordered} 
insulators differs from that of Q1D systems \cite{comment}.

While modeling the full $\gamma_{ab}$ at the critical point needs 
more careful numerical studies, the insulating limit is simpler 
and provides a test case for the generalized DMPK equation (3). It 
predicts that the logarithmic interaction between the transmission 
eigenvalues $\lambda_a$ vanishes as $1/L$ in the insulating limit. 
In this paper we will test this prediction by evaluating the full 
distribution of conductances in the insulating limit for a 3D 
conductor as described by Eq.~(\ref{three}), using the simple 
approximate model for $K$ suggested by our numerical studies 
\cite{noteA1}, namely 
\begin{equation}\label{simple} 
K_{aa}\approx K_{11}\approx 1/2\xi~~{\rm and}~~ \gamma_{ab}\approx 
\gamma_{12}\approx \xi/2L. 
\end{equation} 
We 
can now follow \cite{beenakker} and  map the problem onto a 
Schr\"odinger equation in imaginary time. The corresponding 
Hamiltonian contains an interaction term with strength 
proportional to $\gamma_{12}(\gamma_{12}-2)$ which vanishes not 
only for the Q1D unitary case $\gamma_{12}=\beta=2$ but also in 
the limit $\gamma_{12}\rightarrow 0$. In the present case, 
$\gamma_{12}$ is of order $\xi/2L \ll 1$, and we can neglect the 
interaction and use the $\beta=2$ solution of \cite{beenakker}. 
For $\lambda_a\equiv \sinh^2 x_a$ the 
distribution is then given by 
\cite{mu-wo-99} 
\begin{equation}\label{pg} 
P(g)=\int\prod_a^N dx_a e^{-H}\delta (g-\sum_a h(x_a)); \;\;\; h(x)\equiv 
{\rm sech}^2 x. 
\end{equation} 
Here the $\delta$-function represents 
the Landauer formula for 
the conductance of a multichannel system \cite{landauer}, and the 
`Hamiltonian' $H$ is given, in the insulating limit $x_a \gg 1$, by 
\cite{beenakker} 
\begin{equation}\label{ham} 
H= -\sum_{a>b}^N\left[\frac{1}{2} \ln |\sinh^2 x_a-\sinh^2 
x_b|^{\gamma_{12}}+\ln |x^2_a- x^2_b|\right] - \sum_{a=1}^N 
\left[\frac{1}{2}\ln \sinh 2x_a+\ln x_a-\Gamma x^2_a\right], 
\end{equation} 
where $\Gamma\equiv 1/L_zK_{11}\approx 2\xi/L_z\ll 
1$. The ratio $\Gamma/\gamma_{12}\approx 4 L/L_z$ 
depends on the geometry, but is independent of disorder, leaving a 
single independent parameter $\Gamma$ that determines the strength 
of disorder. We will use $\Gamma/\gamma_{12}=4$ appropriate for a 
cubic system. Note that fluctuations in $k_{aa}$, ignored in the 
model, would make the numerical 
factors in $\Gamma$ and $\gamma_{12}$ inaccurate, but will not change 
either the length or the disorder dependence of these parameters. 
 
The replacement of $\beta=2$ in \cite{beenakker} by 
$\gamma_{12}\rightarrow 0$ 
in Eq.~(\ref{ham}) has the consequence that while all $\langle 
x_a\rangle\gg 1$ in the insulating regime, the difference 
$s=\langle x_{a+1}-x_a\rangle$ is \textsl{not} of the same order 
as $\langle x_a\rangle$. 
For example, if we keep only the first 
two levels, the saddle-point solutions for $x_1$ and $x_2$ give 
$\langle x_1\rangle\sim L_z/\xi$ and $\langle x_2-x_1\rangle\ll 
\langle x_1\rangle$. 
We therefore do not assume that $x_2\gg x_1$. However, we do make 
the simplifying approximation that $ \ln |\sinh^2 x_a-\sinh^2 
x_b|\approx \ln \sinh^2 x_a$ and $\ln |x^2_a- x^2_b|\approx \ln 
x^2_a$ for $a>2$. Eq.~(\ref{ham}) then becomes 
\begin{equation} 
H\approx H_1+\sum_{a=2}^N\left[V(x_a) 
-\gamma_{12}(a-2)f(x_a)\right], 
\end{equation} 
where 
\begin{eqnarray} 
H_1 & = & -  \ln|x^2_2-x^2_1| + \Gamma x^2_1-\frac{1}{2}\ln \sinh 2 
x_1 -\ln x_1,\cr V(x) & = & \Gamma x^2 - \frac{1}{2}\ln \sinh 2x  - \ln 
x - \gamma_{12}\ln \sinh x; \;\;\;  f(x) = \ln \sinh 
x+\frac{2}{\gamma_{12}}\ln x 
\end{eqnarray} 
 
Following ref \cite{mu-wo-99}, we separate out the lowest level 
$x_1$ and treat the rest as a continuum with density $\sigma(x)$ 
beginning at a point $\xx> x_1$. 
The corresponding saddle point 
free energy $F_{sp}(x_1,\xx;g)$ has the form 
\begin{equation} 
F_{sp}(x_1,\xx) =  H_1 - \frac{1}{2\gamma_{12}}\int_{\xx}^b 
\frac{dx}{f'(x)} [V'^2(x)-\mu^2_1 h'^2(x)], 
\end{equation} 
where $\mu_1=-V'(\xx)/h'(\xx)$ and primes denote 
$x$-derivatives. Eq~(\ref{pg}) can then be rewritten as 
\begin{equation}\label{model} 
P(\ln g)\propto g \int_{\xxm}^{b}d\xx 
e^{-F_{sp}(x_1,\xx;g)}e^{-2(\xx-x_1)}, 
\end{equation} 
where the integration over $x_1$ is eliminated by a constraint 
arising from the minimization of the free energy: 
\begin{equation} 
x_1 = \cosh^{-1}[1/\sqrt{g - g_0}]; \;\;\; 
g_0  = -\frac{1}{\gamma_{12}}\int_{\xx}^b 
\frac{dx}{f'(x)}h'(x)[V'(x)+\mu_1h'(x)]. 
\end{equation} 
The lower limit $\xxm$ is the larger of the additional 
constraints imposed by the conditions $\sigma(\xx)\ge 0$ and $\xx 
> x_1 \ge 0$, $x_1 $ real. 
 
\begin{figure}
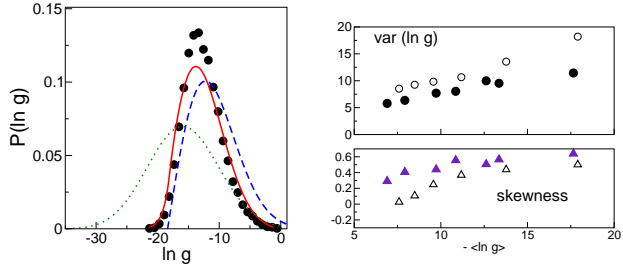
 
\centerline{\includegraphics[clip,width=0.18\textheight]{mmwk-fig4a.eps} 
~~~\includegraphics[clip,width=0.20\textheight]{mmwk-fig4b.eps}} 
\caption{Left: Conductance distribution for 3D insulators obtained 
from direct numerical simulation  for $W=20$, $L=18$ (circles), 
and from Eq.~(\ref{model}) for $\Gamma=0.054$ (solid line). Both have the 
same mean value $\langle\ln 
g\rangle=-12.6$. Dashed and dotted lines show Eq.~(\ref{sp}) with 
$\xxm=1/2\Gamma+5/8$ and $\xxm=1/\Gamma$, respectively, with 
$\Gamma=0.054$. Right: 
Comparison of var($\ln g$) and  skewness 
as a function of $\langle \ln g\rangle$ obtained from numerical 
simulations for various $W$ and $L$ (full symbols) and from 
present model Eq.~(\ref{model}) (open symbols).} 
\label{skewness} 
\end{figure}

Let us consider first a simple approximate solution of 
Eq.~(\ref{model}), which is dominated by the lower limit of the 
integral.  To a good approximation, $g_0$ is negligible compared 
to $g$ in the insulating limit, and $x_1\approx \frac{1}{2}\ln 
(4/g)$. The condition $\sigma(\xx) \ge 0$ gives $\xxm\approx 
(1+\Gamma+\gamma_{12})/2\Gamma$ and hence $F_{sp}\approx H_1$. 
This immediately leads to 
\begin{equation}\label{sp} 
P(\ln g)\propto (4 \xxm^2-u^2) 
e^{-\frac{\Gamma}{4}(\frac{1}{\Gamma}+u)^2}, \;\;\; u\equiv \ln (g/4) 
\end{equation} 
valid for $|u|<2\xxm$. 
Figure \ref{skewness}  shows Eq.~(\ref{sp}) compared with the results from direct integration 
of 
Eq.~(\ref{model}), both compared with numerical results based on Eq.~(\ref{AndHam}). For the 
analytic curves, we chose $\Gamma=0.054$ to have the same 
$<\ln g>$ from Eq.~(\ref{model}) as in the numerical 
case. Note that using the Q1D result $\gamma_{12}=1$ gives 
$\xxm\approx 1/\Gamma$, leading to a log-normal distribution (see 
dotted line in Figure \ref{skewness}). 
A saddle point analysis of Eq.~(\ref{sp}) allows us to obtain $\langle \ln 
g\rangle\approx 1/\Gamma 
-\sqrt{2/\Gamma}$ and var($\ln g$)$\approx 1/\Gamma$, to be compared with 
the corresponding Q1D results $\langle \ln g\rangle\approx 1/\Gamma$ and 
var($\ln g$)$\approx 2/\Gamma$. Moreover the skewness 
$\langle (\ln g- \langle \ln g\rangle)^3\rangle/[\langle (\ln g- \langle 
\ln 
g\rangle)^2]^{3/2}$ saturates to a finite value of order unity independent 
of $\Gamma$ 
for $\Gamma\rightarrow 0$, 
showing the absence of log-normal distribution 
for 3D insulators even at very strong disorder. 
Right figure \ref{skewness} shows variance and skewness calculated from direct integration of 
Eq.~(\ref{model}) compared to numerical results, consistent with saddle point 
results from Eq.~(\ref{sp}). Quantitative differences between Eq.~(\ref{model}) and 
numerical results are due to our 
simplified model Eq.~(\ref{simple}),  which still overestimates the 
strength of the 
interaction for higher channels. 
 
It is instructive to analyze the eigenvalue spectrum in terms of the 
density $\sigma(x)$. In the insulating phase we find 
$\sigma (x)\approx 2\Gamma/\gamma_{12}=8L/L_z$ for $x \gg 2/\gamma_{12}$. 
This corresponds to a uniform average spacing 
$s=\langle x_{a+1}-x_a \rangle$ of eigenvalues of order unity ($L=L_z$), 
compared to the uniform spacing $s\approx L_z/\xi$ in Q1D. In contrast, 3D 
metals are similar to Q1D metals having uniform $\sigma(x)$ extending down 
to $x=0$ and $s\sim L_z/L^2$. The opening of a gap in the spectrum of 
Lyapunov exponents $\nu_n\equiv\langle x_n\rangle/L_z\sim 1/\xi$ 
may be considered as the signature of the Anderson transition.

We conclude that Eq.~(\ref{three}) indeed 
describes a 3D insulator. The solution of Eq.~(\ref{three}), given 
in Eq.~(\ref{model})  proves the non-trivial asymmetry in the 
distribution $P(\ln g)$, which is qualitatively distinct from a 
Q1D insulator. Although such a distribution has been known 
numerically, our method gives for the first time a simple theoretical 
understanding of the entire distribution. 
This opens up the 
possibility to study analytically in more detail 
the insulating regime as well as the 
Anderson transition in 3D in 
terms of the distribution of conductances, providing an 
opportunity to investigate at least qualitatively 
the nature of a quantum phase 
transition in the presence of large mesoscopic fluctuations. 
Numerical work is underway to construct a model for the matrix $K$ 
at the critical point.

\acknowledgements{%
We acknowledge useful discussions with A.D. Mirlin. KAM is 
grateful for support from and hospitality at U. Karlsruhe. PW 
acknowledges support through a Max-Planck Research Award and the 
Center for Functional Nanostructures of the DFG. PM thanks APVT.
}


\begin{thebibliography}{40} 
 
 
 
\bibitem{mu-wo-99} 
\Name{Muttalib K. A. \and W\"olfle P.} 
\REVIEW{Phys. Rev. Lett.}{83}{1999}{3013}; 
\Name{W\"olfle P. \and Muttalib K. A.} 
\REVIEW{Ann. Phys. (Leipzig)}{8}{1999}{753} 
 
 
\bibitem{go-mu-wo-02} 
\Name{Gopar V. A. \etal} 
\REVIEW{Phys. Rev. B}{66}{2002}{174204}; 
\Name{Froufe-Perez L. \etal} 
\REVIEW{Phys. Rev. Lett.}{89}{2002}{246403} 
 
\bibitem{mu-wo-ga-go-03} 
\Name{Muttalib K. A. \etal} 
\REVIEW{Europhys. Lett.}{61}{2003}{95}; 
\Name{Garc\'{\i}a-Martin A.  \and S\'aenz J. J.} 
\REVIEW{Phys. Rev. Lett.}{87}{2001}{116603} 
 
 
\bibitem{markos1} 
\Name{Marko\v{s} P.} 
\REVIEW{Phys. Rev. Lett.}{83}{1999}{588}; 
\Name{R\"uhl\"ander M. \etal} 
\REVIEW{Phys.  Rev. B}{64}{2001}{212202} 
 
\bibitem{markos2} 
\Name{Marko\v{s} P.} 
\REVIEW{Phys. Rev. B}{65}{2002}{104207}; 
\Name{Marko\v{s} P.} \Book{Anderson Transition and its Ramifications} 
\Editor{T. Brandes \and S. Ketteman} \Book{Lecture Notes in Physics} 
\Vol{630} \Publ{Springer} \Year{2003} 
 
\bibitem{lee} For a review, see 
\Name{Lee P. A.\ and Ramakrishnan T. V.} 
\REVIEW{Rev. Mod. Phys.}{57}{1985}{287} 
 
\bibitem{mu-wo-go-03} 
\Name{Muttalib K. A. \etal} 
\REVIEW{Ann. Phys. (NY)}{308}{2003}{156} 
 
\bibitem{comment} In Q1D systems, defined in 
\cite{mu-wo-99,go-mu-wo-02,mu-wo-ga-go-03,dmpk}, disorder is 
always \textsl{weak} enough to assure that  the localization 
length $\xi \gg L$ where $L$ is the transverse dimension. The Q1D 
insulator corresponds to the \textsl{weakly disordered} systems of 
length $L_z \gg \xi$. This is different from localization in 3D 
which occurs at \textsl{strong} disorder, where $\xi \ll$  both 
$L_z$ and $L$. 
 
\bibitem{noteA2}In 
\Name{Altshuler B. L. \etal}
\REVIEW{Sov. Phys. JETP}{64}{1986}{1352} and 
\REVIEW{Phys. Lett. A}{134}{1989}{488}, $P(g)$ has been 
considered in $d=2+\varepsilon$ dimensions where a completely 
log-normal distribution is expected at very large disorder. 
However, this result can not be extended to $d=3$ \cite{mu-wo-99,footn1}. 
 
\bibitem{mu-go-02} 
\Name{Muttalib K. A. \and Gopar V. A.} 
\REVIEW{Phys. Rev. B}{66}{2002}{115318}; 
\Name{Muttalib K. A. \and Klauder J. R.} 
\REVIEW{Phys. Rev. Lett.}{82}{1999}{4272} 
 
\bibitem{muttalib}See e.g. 
\Name{Stone A. D. \etal} 
\Book{Mesoscopic phenomena in solids} 
\Editor{B. L. Altshuler, P. A. Lee \and R. A. Webb} 
\Publ{North-Holland} 
\Year{1991} 
\Page{369} 
 
\bibitem{dmpk} 
\Name{Dorokhov O. N.} 
\REVIEW{JETP Lett.}{36}{1982}{318}; 
\Name{Mello P. A. \etal} 
\REVIEW{Ann. Phys. (N.Y. )}{181}{1988}{290} 
 
\bibitem{note1}The matrix $k_{ab}$ depends on the representation $\alpha$. 
Since 
localization is defined in configuration space, we choose the 
position representation. 
 
 
\bibitem{pendry} 
\Name{Pendry J. B. \etal} 
\REVIEW{Proc. R. Soc. London A}{437}{1992}{67} 
 
\bibitem{pichard} 
\Name{Pichard J.-L. \etal} 
\REVIEW{J. Phys. France}{51}{1990}{587} 
 
\bibitem{footn1} 
\Name{Marko\v{s} P. \and Kramer B.} 
\REVIEW{Philos. Mag. B}{68}{1993}{357} 
 
\bibitem{note} Note that for $|a-b|\gg L$, $\gamma_{ab}$ must decrease as 
$L^{-2}$. From (\ref{two}), $\sum_{b}K_{ab}=1$ so 
the sum of $L^2$ \textit{positive} parameters $\sum_b 
\gamma_{ab}$ is $\propto L^0$ in the insulating regime. We ignore 
this detail in our model. 
 
\bibitem{chalker} 
\Name{Chalker J. T. \and Bernhardt M.} \REVIEW{Phys. Rev. 
Lett.}{70}{1993}{982} 
 
 
\bibitem{noteA1}We only consider orthogonal symmetry. As suggested 
in 
\Name{Lerner I. V. \and Imry Y.}
\REVIEW{Europhys. Lett.}{29}{1995}{49}  and 
\Name{Kettemann S. \and Raikh M. E}
\REVIEW{Phys. Rev. Lett.}{90}{2003}{146601}, we 
expect non-trivial dependence of the localization length $\xi$ on 
the universality class in 3D.
 
 
\bibitem{beenakker} 
\Name{Beenakker C. W. J. \and Rejaei B.} 
\REVIEW{Phys. Rev. Lett.}{71}{1993}{3689}; 
\REVIEW{Phys. Rev. B}{49}{1994}{7499} 
 
 
 
\bibitem{landauer} 
\Name{Landauer R.} 
\REVIEW{IBM J. Res. Dev.}{1}{1957}{223} 
 
 
\end{thebibliography}
\end{document}